\documentclass[preprint2]{aastex}

\newcommand{\uzfor}{\mbox{UZ~For}}

\newcommand{\xmm}{\mbox{\em XMM-Newton}}
\newcommand{\exosat}{\mbox{\em EXOSAT}}
\newcommand{\rosat}{\mbox{\em ROSAT}}
\newcommand{\euve}{\mbox{\em EUVE}}

\newcommand{\dex}[1]{\hbox{$\times\hbox{10}^{#1}$}}
\newcommand{\kev}{\,\mbox{keV}}
\newcommand{\msun}{\,\mbox{$\mbox{M}_{\odot}$}}

\newcommand{\etal}{\mbox{et\ al.\ }}

\slugcomment{Accepted for publication in Astrophysical Journal Letters}

\shorttitle{A burst from UZ~For}
\shortauthors{M.~Still and K.~Mukai}

\begin{document}

\title{A burst from the direction of UZ Fornacis with {\em XMM-Newton}}

\author{Martin Still\altaffilmark{1} and Koji Mukai\altaffilmark{1}}
\affil{NASA/Goddard Space Flight Center, Code 662, Greenbelt, MD~20771}

\altaffiltext{1}{Universities Space Research Association}

\begin{abstract} 

The \xmm\ pointing towards  the magnetic cataclysmic variable  \uzfor\
finds   the source to  be   a factor $>   10^3$  fainter than previous
\exosat\  and \rosat\ observations.  The  source  was not detected for
the majority of a  22 ksec exposure  with the EPIC cameras, suggesting
that  the  accretion  rate either   decreased, or stopped  altogether.
However a  1.1   ksec burst was    detected from \uzfor\    during the
observation.   Spectral    fits  favour    optically  thin,   $kT    =
4.4^{+7.6}_{-1.8}$ keV  thermal emission.  Detection   of the burst by
the on-board Optical Monitor indicates that  this was most probably an
accretion event.  The 0.1--10 keV luminosity of $2.1^{+0.6}_{-0.3}$
\dex{30} erg s$^{-1}$  is  typical for accretion shock  emission  from
high state polars  and would result from  the potential energy release
of $\sim  10^{16}$ g of gas.  There  is no significant soft excess due
to reprocessing in the white dwarf atmosphere.
\end{abstract}


\section{Introduction}
\label{sec:introduction}

Polars  contain a white  dwarf   with a rotation period  synchronously
locked  to a   binary  orbit and   which accrete magnetically-threaded
material from a Roche lobe  filling companion star (Kube, G\"{a}nsicke
and Beuermann 2000).    Long-term  monitoring has revealed  that  many
polars undergo  a  cycle of   high- and low-states.   High states  are
generally characterized  by  thermal  emission in X-ray  bands  (Done,
Osborne  \& Beardmore 1995), with  cyclotron and illuminated companion
star emission contributing  at lower  energies (Ferrario \etal  1989).
In the lowest  states,  X-rays  are not  detected and  both  intrinsic
emission from the companion star and  a Zeeman spectrum from the white
dwarf dominate, indicating  that accretion has almost ceased (Ferrario
\etal 1992).  An \xmm\ pointing finds the polar \uzfor\ in a low X-ray
state, except during   a burst of duration   1.1  ksec. This  type  of
behaviour has  not previously been seen  during X-ray pointings of low
state polars, and we investigate whether the burst was the result of a
discrete accretion event.

\begin{figure*}
\begin{picture}(100,0)(10,20) 
\put(0,0){\includegraphics{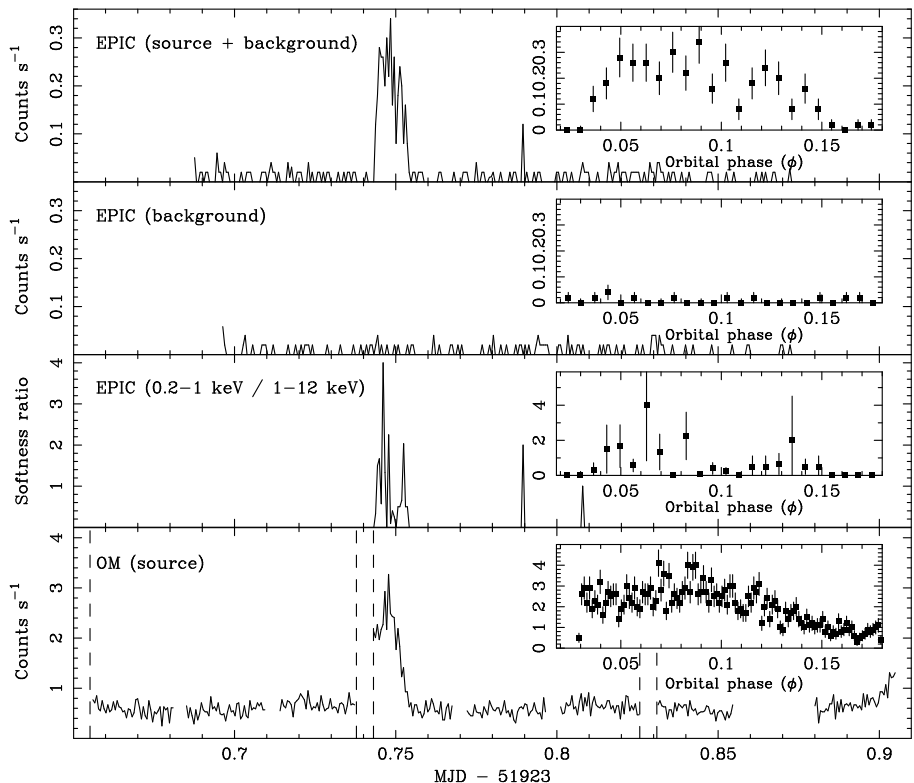}}
\noindent
\end{picture}
\vspace{80mm}
\figcaption[f1.ps]{The top panel provides the MOS+pn source and 
background count  rate, binned into intervals  of  50 sec.  The second
panel is  the MOS+pn background   count rate and  the  third panel the
softness ratio.  The bottom  panel displays  the background-subtracted
OM source  count  rate.  The insets  provide a  blow-up  of the burst,
binned  over the orbital  ephemeris of  Perryman \etal  2001.   The OM
inset is sampled at 10 sec. Dashed lines correspond to eclipse ingress
and egress.\label{fig:lc}}
\end{figure*}

After its detection in the \exosat\ archive (EXO~033319-2554.2; Giommi
\etal 1987), UZ For  was identified as an eclipsing  polar with  a 127
min orbital period (Osborne \etal  1988; Beuermann, Thomas and Schwope
1988).  During high states, gas accretes onto  a small fraction of the
white dwarf surface through standing shocks above  one or more surface
pole  caps, where the  field strength  has been measured  at  53 MG by
Rousseau \etal (1996).   Shocked gas is  heated  to $\sim 10^8$~K  and
subsequently cooled  by  hard  Bremsstrahlung (Beuermann,   Thomas and
Pietsch 1991), cyclotron   emission   (Berriman and Smith    1988) and
Compton scattering (Done and Magdziarz  1998).  Polars are  detectable
as soft  blackbody sources either from hard  X-rays thermalized by the
surface  of   the white  dwarf (Ramsay  \etal   1993; see also Ramsay,
Cropper  and Mason  1996),  or diffused  through  the atmosphere after
dense bullets   have penetrated below  the photosphere   (Kuijpers and
Pringle  1982).  The diffusion  model is preferred for \uzfor\ because
the combined Bremsstrahlung and cyclotron luminosity is not sufficient
to power the soft X-ray  component through thermalization (Frank \etal
1988; Ramsay   \etal  1993).   A  thorough review   of  magnetospheric
interaction,  accretion  dynamics and  emission  processes is given by
Warner (1995).

\section{Observations}
\label{sec:observations}

\xmm\  (Jansen \etal  2001) observed  \uzfor\  during the  Performance 
Verification  phase of telescope  commissioning  on 2001  Jan 14.  The
EPIC  MOS,  EPIC pn, RGS and   OM detectors were  all  open during the
pointing.  The EPIC CCDs were configured in Large Window Mode and were
integrated over 19 ksec (MOS1 and MOS2) and 22 ksec (pn).  EPIC events
were re-reduced using the  standard pipelines within the \xmm\ Science
Analysis System (SAS) version  5.1, using the Calibration Access Layer
current  at   2001 July  3.  Both  RGS   detectors were  configured in
spectroscopy mode but no detection of \uzfor\ was found in reduced RGS
events over  integrations of 22 ksec.   Eight consecutive exposures of
2.2 ksec were made by  the OM in timing  mode with the 240--360nm UVW1
filter.

\subsection{Light curve}
\label{sub:lc}

To reduce background from both MOS and pn  cameras, single events were
selected with corrected pulse heights     $> 200$ eV from within     a
circular  aperture of radius  23    arcsec, centered on  the   source.
Background events were  selected from an   aperture of the  same size,
centered at an  arbitrary source-free region, on the  same chip as the
target.  Events from all three  cameras were combined and total source
and background light curves are  presented in Fig.~\ref{fig:lc} with a
sampling  of 50 sec.  Event arrival times have   been corrected to the
solar system barycenter.

For a large fraction of the pointing, the  count rates from within the
aperture  are consistent with diffuse background.    From the model of
Ramsay \etal  (1993;   based  on  \rosat\   high-state  observations),
consisting of a  moderately   absorbed 28 eV  blackbody  with  a  hard
Bremsstrahlung tail,  we expect a total  of 3.1 events per second from
the EPIC  detectors.  Unless accretion  emission was  arriving outside
the EPIC  bands (i.e., the  accretion temperature was cooler), \uzfor\
was caught  during a deep  low  state.  Given  the observed background
rate, a 3-$\sigma$  detection would require  $10^{-3}$ counts s$^{-1}$
in   the aperture.  Therefore  the  X-ray  flux  is generally  $>$ 2.8
\dex{3} times fainter than during the \rosat\ pointing in the 0.1--2.0
keV band.

However, for a duration  of 1.1 ksec,  centered at MJD~51923.75, \xmm\
detected  a burst from the  direction of \uzfor.  A sliding-box source
detection   algorithm   provides   2000  sky  coordinates   of   RA  =
$03^{\mbox{\tiny h}}  35^{\mbox{\tiny m}} 28^{\mbox{\tiny s}}.69(4) $,
$\delta =   25^\circ 44^\prime  23^{\prime\prime}.4(6)$,  in excellent
agreement  with the optical counterpart  of \uzfor\  (Downes and Shara
1993).  The bracketed numbers are 1-$\sigma$ uncertainties on the last
digit.  Time-dependent  softness   ratios were constructed  using  the
bandpassses 0.2--1 keV and 1--12 keV.  The ratios were  set to zero in
time  bins  unpopulated   by hard events.    Softness  behaviour (also
presented  in  Fig.~\ref{fig:lc})    is constrained  by     low-number
statistics, but  gives the impression that  the rise and  decay of the
burst are characterized  by a soft  spectrum which hardens during  the
peak of the event.

\begin{figure*}                           
\begin{picture}(100,0)(10,20)
\put(0,0){\includegraphics{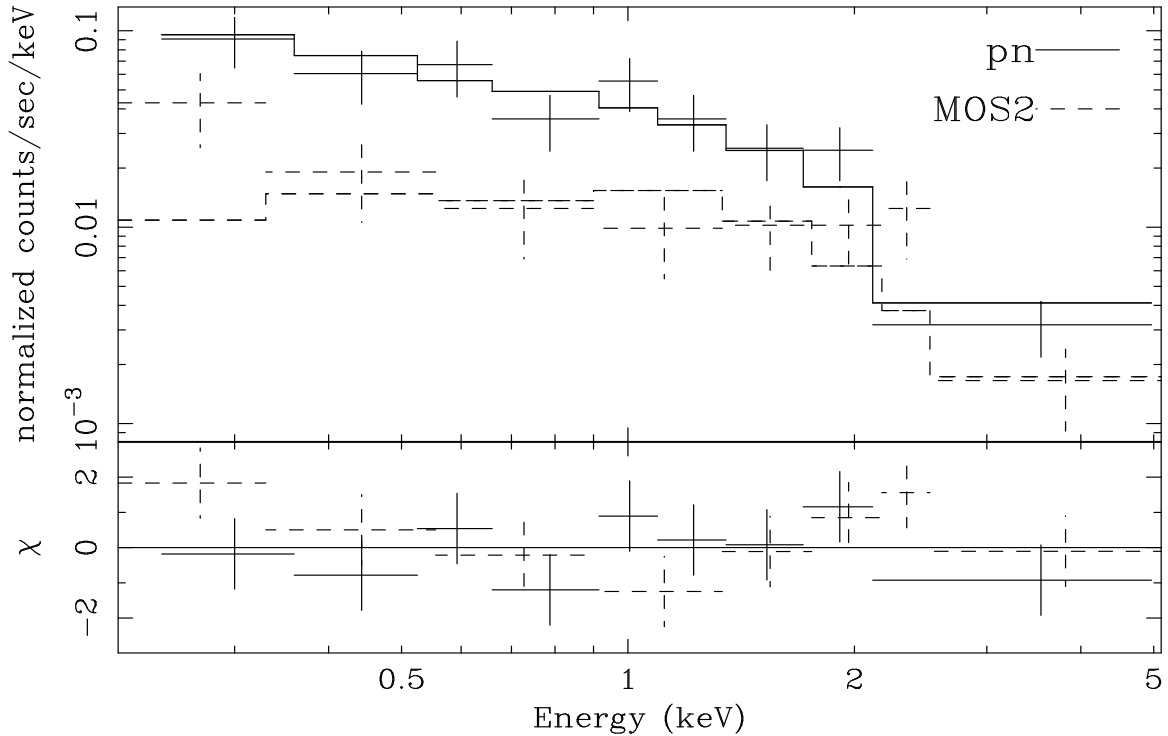}}     
\noindent    
\end{picture}    
\vspace{90mm}
\figcaption[f2.ps]{Energy  spectra for the  EPIC  MOS2 and pn cameras.
The MOS1 spectrum  has  not been  included for   presentation clarity.
Energy channels  have been  binned  so  that signal-to-noise  in  each
channel is $> 5$ (MOS2) and $> 10$ (pn).\label{fig:spectrum}}
\end{figure*}

The UV  counterpart to the    burst was detected   by  the OM over   a
relatively constant number of source counts, probably intrinsic to the
companion star. Since the OM timing analysis software is not currently
available, OM events were    extracted from the event   files  without
pointing  corrections or grade  selection.   The start times for  each
exposure  were taken from the  Non-Periodic  Housekeeping file with an
uncertainty of 1 s. Source and background apertures were both circular
regions of radius 7 detector pixels. Note  that the OM counts increase
by a factor 2 at the end of the pointing, after  the EPIC cameras have
been turned off, perhaps indicating the onset of another event.

\subsection{Energy spectrum}
\label{sub:spectrum}

Ramsay \etal (1993) found a small hard X-ray excess  in a \rosat\ PSPC
pointing of \uzfor\ which they interpreted as Bremsstrahlung emission,
however the majority of flux was confined to the soft energy bands and
fit acceptably by  a blackbody of  $\sim$  28 eV.  In  this section we
will attempt to fit similar models to the burst spectrum. We note that
the softness  ratio from Fig.~\ref{fig:lc} possibly indicates spectral
variability  during the  burst,  but  small-number  photon  statistics
prevent us from conducting time-dependent spectral fitting.

Single source  events  with pulse  heights $>   100$ eV  were selected
within  the time  range MJD 51\,923.7435--51\,923.7562,  from the same
sky region as    in  Sec.~\ref{sub:lc}.  Background count   rates  are
1.2\dex{-3} counts s$^{-1}$ in the  MOS cameras and 4.8\dex{-3} counts
s$^{-1}$ in  the pn  camera  within the  defined filter region,  i.e.,
there are $\sim$  2 and 6  background events contaminating  the source
spectrum of each MOS and the pn cameras, respectively.  Since both the
source and background are subject to  low-number statistics, we do not
subtract a background spectrum.

Events were  binned spectrally to provide  a signal-to-noise $> 5$ and
10 in each channel of the  MOS and pn  detectors, respectively.  MOS1,
MOS2 and pn spectra were folded through appropriate response files and
fit together  using  a maximum-likelihood  scheme (Cash 1979)   with a
single  blackbody model, absorbed by  a neutral column.  A C-statistic
of   42.3    over  22    spectral   bins  indicates  a  blackbody   of
$440^{+60}_{-50}$ eV with a  flux of $2.9^{+0.5}_{-0.4} \dex{-13}$ erg
s$^{-1}$ cm$^{-2}$ and negligible  neutral absorption.  This is hotter
than expected from thermal emission on  the white dwarf surface (e.g.,
Ramsay  \etal  1993).  A  Monte Carlo  test, using  1000 realizations,
created   fake  spectra based  on  the  best model parameters  and the
observed counts.  This successfully produced an  improved fit over the
data for 99.7 percent of the realizations.  Therefore the model is not
a good fit statistically.

With    a    C-statistic  of  21.8    over      22 spectral bins,    a
$4.4^{+7.6}_{-1.8}$  keV Bremsstrahlung  model,  absorbed by a neutral
column  of  $3.2^{+2.5}_{-1.8}    \dex{20}$  cm$^{-2}$,   provides  an
acceptable fit. The Monte Carlo test provided an improved fit over the
data with  48.8 percent  of 1000  realizations.  The  neutral hydrogen
column density through  the galaxy in  the  direction of  \uzfor\ is 1
\dex{20} cm$^{-2}$ (Dickey and Lockman 1990).  The emission measure is
given by  $\int n_e^2  \mbox{d}V$  = $4.3^{+1.3}_{-0.6}  \dex{11} D^2$
cm$^{-3}$, where $D$  is the source  distance, $n_e$ the fully-ionized
electron density, and $V$ the emitting volume.  Data and fits from the
MOS2 and pn cameras are presented in Fig.~\ref{fig:spectrum}.

Although the MOS2 data show a possible  excess in the softest channel,
a fit  of absorbed  Bremsstrahlung and blackbody  components, combined
over all three EPIC detectors,  results  in a blackbody of  negligible
intensity,  yielding no improvement  in fit quality.   Fits of greater
sophistication (e.g., multi-temperature  shocks  and  Compton cooling;
Ramsay \etal 2001) are not justified by the data.

\section{Discussion}
\label{sec:discussion}

Two similar bursts, of  duration $\sim 1$  hr, were observed during an
otherwise  non-detection  of the  polar QS~Tel  using \euve\ by Warren
\etal (1993).  Using the family of solutions that Warren \etal\ were
able to constrain,   the Bremsstrahlung emission observed   during the
\uzfor\ burst is at  least  an order of   magnitude fainter  than  the
QS~Tel events.

Adopting a distance  to \uzfor\  of $208  \pm  40$ pc (Ferrario  \etal
1989), the emission measure is $8.7^{+2.6}_{-1.2}$ \dex{52} cm$^{-3}$.
This is consistent with flares from rapidly rotating M dwarfs, similar
to  the  companion star,  which show   typical  emission  measures  of
$10^{52\mbox{-}54}$  cm$^{-3}$ (e.g., Pan and   Jordan 1995; Pan \etal
1997).  The derived  electron  temperature is also typical  of coronal
activity on  late-type stars ($\sim  10^7$  K; Pan and  Jordan  1995).
Stellar flares are generally characterized  in two classes which, like
the current burst, soften spectrally at the beginning  and end of each
event  (Tsikoudi  and Kellett 2000).    Long-decay flares have similar
soft spectra to the \uzfor\ event, but with decline times of the order
a few hours.  Impulsive flares have rapid rise and decay times similar
to the present burst, but generally  have shorter lifetimes and harder
spectra ($> 20$ keV;  Schmitt, Haisch and  Barwig 1993).  The  \uzfor\
burst fits into neither category.

However, the UV detection of the burst indicates the event is unlikely
to be coronal in origin, but UV emission  is a predictable consequence
of  white dwarf  accretion (e.g.\ Warner  1995).  Without simultaneous
low-energy  observations, the nature of the  burst would have remained
ambiguous.  This event illustrates the advantages of flying optical/UV
detectors on-board high energy missions.  Again assuming a distance of
208 pc, the 0.1--10   \kev\  burst luminosity  is  $2.1^{+0.6}_{-0.3}$
\dex{30} erg s$^{-1}$.  This is consistent with the typical hard X-ray
luminosities of  high  state polars (Chanmugam,  Ray  and Singh 1991).
Assuming that all  the accretion energy is emitted  in the EPIC bands,
then    $L_{\mbox{\tiny   X}}     =     G     \dot{M}   M_{\mbox{\tiny
WD}}/R_{\mbox{\tiny   WD}}$.  $L_{\mbox{\tiny   X}}$    is the   X-ray
luminosity,   $G$  the  gravitational constant,    $\dot{M}$  the mass
accretion   rate, $M_{\mbox{\tiny WD}}$   the   white dwarf mass   and
$R_{\mbox{\tiny WD}}$  the  white dwarf radius.  If   we consider a  1
solar mass white dwarf of radius $10^9$ cm, the mass accretion rate is
1.6 \dex{13}  g s$^{-1}$.  1.7 \dex{16}  g of  material will have been
accreted during the burst.

We find  no   statistically-significant evidence for a   soft spectral
component    due to   white  dwarf   reprocessing.   Consequently, the
accretion density   must be relatively   low in order   to prevent gas
piercing  below   the white  dwarf  photosphere  (Frank   \etal 1988).
However   illumination from above  by  the  shock  should result in  a
photospheric  blackbody  temperature of $\sim$  20 eV,  which would be
detectable in  the   softest  channels of  the  EPIC  detectors (e.g.,
Beuermann, Thomas and Pietsch 1991).  Similarly, \xmm\ observations of
WW~Hor in an intermediate accretion state  find no evidence for a soft
component.   (Ramsay  \etal   2001) argue  that   this   could be  the
consequence  of   a larger-than-usual  accretion   zone, resulting  in
reprocessed emission too cool to be detected with the EPIC cameras.

To check the visibility of the  major accreting pole on \uzfor\ during
this burst,  we  fold the  data over two  orbital ephemerides measured
from  optical eclipse timing.  The  ephemeris of Perryman \etal (2001)
indicates that the burst begins at $\phi$ = 0.03--0.04, where $\phi$ =
0.000 is defined as superior conjunction  of the white dwarf.  At this
phase the main high-state accretion spot is on the visible side of the
white  dwarf.  However this  also corresponds  to  the egress phase of
white dwarf eclipse which occurs rapidly over 1--3 s in optical bands.
The characteristic scatter  of eclipse times  about this ephemeris  is
large,  $\sim$ 50 s,   possibly  indicating systematics  caused  by  a
migrating accretion spot.  However it is  plausible that the perceived
start of the burst  is in fact  eclipse egress, and the burst started,
in  reality, during white dwarf  eclipse.  Consequently, the intrinsic
burst may  have been up  to 450 s longer  than observed.   From an eye
inspection of the  EPIC  event list, the  first  detected X-ray photon
that   may be    associated     with  the  burst    was   received  at
MJD~51923.7438. Since this is $8  \dex{-3}$ of an orbital cycle  after
eclipse egress, there is no evidence that the burst was visible during
eclipse.  The  X-ray and UV bursts,   binned on the  orbital phase are
presented in the insets  to Fig.~\ref{fig:lc}.  The  rise to  UV burst
occurred within  a 10 s   period, consistent with the optical  eclipse
egress.   Although statistically the   X-ray   and UV bursts   started
simultaneously, there is   some suggestion (albeit  based on low-count
statistics) that the X-ray burst takes somewhat longer to rise.

Unfortunately, the OM data gaps coincide with each white dwarf eclipse
(Fig.~\ref{fig:lc}).  Therefore there   are no spatial constraints  on
the source of  the low-state UV  flux.  Extrapolation of the models of
Houdebine \etal (1996) indicate that the luminosity from an M dwarf of
radius 0.2 \msun\ could be as large as  2 \dex{29} erg s$^{-1}$ in the
UVW1  bandpass, although there is some  ambiguity due to the nature of
chromospheric structure.  This predicts  a maximum companion star flux
of  4 \dex{-17}  erg  s$^{-1}$ cm$^{-2}$  \AA$^{-1}$.   Neglecting CCD
readout deadtimes (which are currently unavailable), the OM count rate
indicates  a   source  flux  of  3 \dex{-16}    erg s$^{-1}$ cm$^{-2}$
\AA$^{-1}$.  Therefore, in the absence of an accretion stream, a large
fraction  of the low-state   UV flux must  have  a source on the white
dwarf, where the flux is reasonable for a white dwarf of radius $10^9$
cm, emitting like a blackbody at 20\,000 K.  At  its peak, the UV flux
in the  burst is   1.4  \dex{-15} erg s$^{-1}$   cm$^{-2}$ \AA$^{-1}$,
neglecting deadtime. With a surface magnetic field of 53 MG, we do not
expect  cyclotron emission  in   the UVW1 passband.   Therefore  it is
likely the UV burst occurs in the heated polar cap  of the white dwarf
(c.f.\ Gaensicke \etal 1998).

\section{Conclusion} 
\label{sec:conclusion}

During  a  low accretion state of   the  magnetic cataclysmic variable
\uzfor, \xmm\ detected a burst with a duration of 1.1 ksec.  The event
is best characterized by an absorbed Bremsstrahlung spectrum, with the
temperature possibly   directly  correlated   with   intensity.   This
suggests that the  emission region is either  a coronal flare from the
companion star,  or a standing  shock above the   magnetic pole of the
white dwarf.  The temperature, emission  measure and colour  evolution
of  the burst   are  consistent with   a stellar   flare, however  the
brightness variability and spectral shape do not fit conveniently into
either class of long-decay   or impulsive flares.   Detection of  a UV
counterpart to the  X-ray burst  strongly   suggests that this is   an
accretion event, where the 0.1--10 keV luminosity is of the same order
as the mean Bremsstrahlung luminosity of high state polars.  The burst
timing suggests that it  is  associated with  the pole  that dominates
accretion during high states and that the event  may have begun during
white dwarf eclipse.

This  behaviour has not been  found previously in polars, although the
EUVE counterparts of similar events  may have been detected by  Warren
\etal  (1993).   As  a conjecture,  this  burst  may be  indicative of
occasional, but significant, amounts  of  companion star gas   falling
through the $L_1$ point of the binary  during low accretion states. If
similar events were detected in the future, it would be interesting to
correlate this activity with  the  long term high/low state  accretion
cycle.  Perhaps  the  frequency  of  these events   indicate impending
switches from the low to high state.

\acknowledgments  
Based on observations  obtained  with {\em XMM-\linebreak Newton},  an
ESA science mission with instruments and contributions directly funded
by ESA  Member States and  the USA (NASA).  We  thank Bing Chen, Steve
Drake, Fred Jansen and the referee for valuable assistance.

\newpage\clearpage


\begin{thebibliography}{}

\bibitem[]{ber88} Berriman G., Smith P. S., 1988, ApJ, 329 L97

\bibitem[]{beu87} Beuermann K.,  Thomas H.-C., Schwope  A, 1988, A\&A,
195, l15

\bibitem[]{beu91} Beuermann K., Thomas H.-C., Pietsch W., 1991, A\&A, 
246, L36 

\bibitem[]{cas79} Cash W., 1979 ApJ, 228, 939

\bibitem[]{cha91} Chanmugam G., Ray A., Singh K. P., 1991, ApJ, 375, 600

\bibitem[]{dic90} Dickey J. M., Lockman F. J., 1990, ARAA, 28, 215

\bibitem[]{don95} Done C., Osborne J. P., Beardmore A. P., 1995, MNRAS,
276, 483

\bibitem[]{don98} Done C., Magdziarz P., 1998, MNRAS, 298, 737

\bibitem[]{dow93} Downes R. A., Shara M. M., 1993, PASP, 105, 127

\bibitem[]{fer89} Ferrario L., Wickramasinghe D.  T., Bailey J., Tuohy
I. R., Hough J. H., 1989, ApJ, 337, 832

\bibitem[]{fer92} Ferrario L., Wickramasinghe D. T., Bailey J., Hough 
J. H., Tuohy I. R., 1992, MNRAS, 256, 252

\bibitem[]{fra88} Frank J., King A. R., Lasota J.-P., 1988, A\&A, 193, 
113

\bibitem[]{gae98} Gaensicke B. T., Hoard D. W., Beuermann K., Sion
E. M., Szkody P., 1998, A\&A, 338, 933

\bibitem[]{gio87} Giommi  P.,   Angelini L., Osborne   J.,  Stella L.,
Tagliaferri G., Beuermann K., Thomas H.-C., 1987, IAU Circ. 4486

\bibitem[]{hou96} Houdebine E. R., Mathioudakis M., Doyle J. G., Foing
B. H., 1996, A\&A, 305, 209

\bibitem[]{jan01} Jansen F. et~al., 2001, A\&A, 365, L1

\bibitem[]{kub00}  Kube J., G\"{a}nsicke   B. T., Beuermann K.,  2000,
A\&A, 356, 490

\bibitem[]{kui82} Kuijpers J., Pringle J. E., 1982, A\&A, 114, L4

\bibitem[]{osb88} Osborne J.  P., Giommi P., Angelini L.,  Tagliaferri
G., Stella L., 1988, ApJ, 328, L45

\bibitem[]{pan97} Pan H. C., Jordan C., Makishima K., Stern R. A., 
Hayashida K., Inda-Koide M., 1997, MNRAS, 285, 735

\bibitem[]{pan95} Pan H. C., Jordan C., 1995, MNRAS, 272, 11

\bibitem[]{per01} Perryman M. A. C., Cropper M., Ramsay G., Favata F.,
Peacock A., Rando N., Reynolds A., 2001, MNRAS, 324, 899

\bibitem[]{ram93} Ramsay G., Rosen S. R.,  Mason K. O., Cropper M. S.,
Watson M. G., 1993, MNRAS, 262, 993

\bibitem[]{ram96} Ramsay  G.,  Cropper M., Mason  K.  O., 1996, MNRAS,
278, 285

\bibitem[]{ram01} Ramsay G., Cropper  M.,  C\'{o}rdova F., Mason   K.,
Much R., Pandel D., Shirey R., 2001, MNRAS, in press

\bibitem[]{rou96}  Rousseau Th., Fischer  A.,  Beuermann K., Woelk U.,
1996, A\&A, 310, 526

\bibitem[]{scm93} Schmitt J. H. M. M., Haisch B., Barwig H., 1993, ApJ,
419, L81

\bibitem[]{tsi00} Tsikoudi V., Kellett B. J., 2000, MNRAS, 319, 1147

\bibitem[]{war93}  Warren J. K., Vallerga  J.  V., Mauche C. W., Mukai
K., Siegmund O. H. W., 1993, ApJ, 414, L69

\bibitem[]{war96} Warner B., 1995, in Cataclysmic Variable Stars
(Cambridge: Cambridge Univ. Press), 307

\end{thebibliography}
\end{document}